\documentclass[11pt]{article}
\usepackage{geometry} 
\usepackage[english]{babel}
\usepackage[english]{layout}

\usepackage{mathpazo} 
\usepackage[scaled]{helvet} 
\usepackage{courier} 
\normalfont
\usepackage[T1]{fontenc}

\usepackage{array} 
\usepackage{paralist} 
\usepackage{verbatim} 

\usepackage[utf8]{inputenc} 
\usepackage{textcomp} 
\usepackage{graphicx}  
\usepackage{epstopdf} 
\usepackage{flafter}  

\usepackage{amsfonts,latexsym,amsthm,amssymb}
\usepackage{amsmath,amssymb,bbm,amsfonts}  
\usepackage{bm}  

\usepackage{memhfixc}  
\usepackage{fancyhdr} 
\usepackage{sectsty}
\allsectionsfont{\sffamily\mdseries\upshape} 

\numberwithin{equation}{section}

\def\eb{\begin{eqnarray*}}
\def\ee{\end{eqnarray*}}

\begin{document}

\title{\LARGE \textbf{Jordanian deformation of the open $s\ell(2)$ Gaudin model}}

\author{\textsf{N. ~Cirilo ~Ant\'onio,}
\thanks{E-mail address: nantonio@math.ist.utl.pt}
\textsf{ ~ N. ~Manojlovi\'c}
\thanks{E-mail address: nmanoj@ualg.pt}
\textsf{ and Z. ~Nagy}
\thanks{E-mail address:
zoltan.nagy@m4x.org} \\
\\
\\
\textit{$^{\ast}$Centro de An\'alise Funcional e Aplica\c{c}\~oes}\\
\textit{Instituto Superior T\'ecnico, Universidade T\'ecnica de Lisboa} \\
\textit{Av. Rovisco Pais, 1049-001 Lisboa, Portugal} \\
\\
\textit{$^{\dag\ddag}$Grupo de F\'{\i}sica Matem\'atica da Universidade de Lisboa} \\
\textit{Av. Prof. Gama Pinto 2, PT-1649-003 Lisboa, Portugal} \\
\\
\textit{$^{\dag}$Departamento de Matem\'atica, F. C. T.,
Universidade do Algarve}\\
\textit{Campus de Gambelas, PT-8005-139 Faro, Portugal}\\
}
\date{}


\maketitle
\thispagestyle{empty}
\vspace{10mm}
\begin{abstract}
We derive the deformed $s\ell(2)$ Gaudin model with integrable boundaries. Starting from the Jordanian deformation of the $SL(2)$-invariant Yang R-matrix and generic solutions of the associated reflection equation and the dual reflection equation,
the corresponding inhomogeneous spin-\textonehalf \ XXX chain is obtained. The quasi-classical expansion of the transfer matrix yields the deformed $s\ell(2)$ Gaudin Hamiltonians with boundary terms.
\end{abstract}

\clearpage
\newpage

\section{Introduction}

Gaudin models have applications in many areas of modern physics, from quantum optics \cite{Garraway11,BogoliubovKulish12} to physics of metallic nano-grains, see \cite{AmicoOsterloh12} and reference therein.
A model of interacting spins in a chain was first considered by Gaudin \cite{Gaudin76,Gaudin83}. In his approach, these models were introduced as a quasi-classical limit of the integrable quantum chains. Moreover, the Gaudin models were extended to any simple Lie algebra, with arbitrary irreducible representation at each site of the chain \cite{Gaudin83}.

The rational $s\ell(2)$ invariant model was studied in the framework of the quantum inverse scattering method \cite{Sklyanin89}.  In his studies, Sklyanin used the $s\ell(2)$ invariant classical r-matrix. A generalization of these results to all cases when skew-symmetric r-matrix satisfies the classical Yang-Baxter equation \cite{BelavinDrinfeld} was relatively straightforward \cite{Semenov97}. Therefore, considerable attention has been devoted to Gaudin models corresponding to the the classical r-matrices of simple Lie algebras \cite{Jurco90} and Lie superalgebras \cite{KulishManojlovic01, KulishManojlovic03,KurakLima04}.  In the case of the $s\ell(2)$ Gaudin system, its relation to Knizhnik-Zamolodchikov equation of conformal field theory \cite{BabujianFlume,FeiginFrenkelReshetikhin,ReshetikhinVarchenko} or the method of Gauss factorization \cite{Sklyanin99}, provided alternative approaches to computation of correlation functions. The non-unitary r-matrices and the corresponding Gaudin models have been studied recently, see \cite{Skrypnyk09} and the references therein.

The quantum inverse scattering method (QISM) \cite{TakhFadI,KulishSklyanin82,Faddeev} as an approach to construct and solve quantum integrable systems has lead to the theory of quantum groups \cite{Drinfeld, Jimbo85}. A particularly interesting feature of quantum groups is a transformation that is called twist \cite{Drinfeld90} and allows constructing new quantum groups from already known ones. Although the twist transformations generate an equivalence relation between quantum groups, they produce different R-matrices. These new R-matrices can in turn lead to new integrable systems \cite{Kulish09}.

Twist of a quantum group, or more general Hopf algebra $\mathcal{A}$, is a similarity transformation of the coproduct $\Delta: \mathcal{A}\rightarrow \mathcal{A} \otimes \mathcal{A}$ by an invertible twist element
$$\mathcal{F} = \sum _{j} f_j^{(1)}\otimes f_j^{(2)} \in \mathcal{A}\otimes \mathcal{A},$$
\begin{equation}
\label{twDelta}
\Delta (a) \to \Delta_{t} (a) = \mathcal{F} \Delta (a) \mathcal{F}^{-1}, \quad a \in \mathcal{A}.
\end{equation}
In order to guarantee the coassociativity property of the coproduct, the element $\mathcal{F}$ has to satisfy certain compatibility condition, the so-called twist equation \cite{Drinfeld90}
\begin{equation}
\label{TwEq}
\mathcal{F}_{12} \left( \Delta \otimes \mathrm{id} \right) (\mathcal{F}) = \mathcal{F}_{23}\left(\mathrm{id} \otimes  \Delta\right) (\mathcal{F}) ,
\end{equation}
where
$$( \Delta \otimes \mathrm{id} ) \sum _{j} f_j^{(1)} \otimes f_j^{(2)}= \sum _{j} \Delta(f_j^{(1)}) \otimes f_j^{(2)} \in \mathcal{A}\otimes \mathcal{A}\otimes \mathcal{A}.$$
The transformation law of the coproduct also determines how the corresponding universal $R$-matrix is changed
\begin{equation}
\label{twtR}
\mathcal{R} \to \mathcal{R} ^{(t)} = \mathcal{F}_{21} \mathcal{R} \mathcal{F}^{-1}, \quad \mathcal{F}_{21} = \sum _{j} f_j^{(2)} \otimes  f_j^{(1)}.
\end{equation}
This new $R$-matrix allows us to build and study new integrable models \cite{Kulish09}.

A particular solution of the twist equation is provided by the Jordanian twist element for the universal enveloping algebra $U(s\ell(2))$ of the $s\ell(2)$ Lie algebra \cite{Gerstenhaber,Ogievetsky}. It was extended to the $s\ell(n)$ case in \cite{KulishLyakhovskyMudrov,Kulish99}. The Yang R-matrix determines the Yangian $\mathcal{Y}(s\ell(n))$, see \cite{CharyPressley}. The Heisenberg XXX spin chain is related to the Yangian $\mathcal{Y}(s\ell(2))$ and the universal enveloping algebra of $s\ell(2)$ is a Hopf subalgebra of the Yangian, $U(s\ell(2)) \subset \mathcal{Y}(s\ell(2))$.  It can be shown that the transformation of the Yang R-matrix by the Jordanian twist element yields the R-matrix of the twisted Yangian $\mathcal{Y}_\theta(s\ell(2))$ \cite{KulishStolin97,KhoroshkinStolinTolstoy}. Because the twist preserves the regularity of the R-matrix the Hamiltonian of the deformed  Heisenberg XXX spin chain with periodic boundary conditions can be calculated \cite{KulishStolin97}. Although it can be seen that the extra terms added to the XXX Hamiltonian do not change the spectrum of the model, the explicit form of the Bethe states is not obvious.

The deformation by the Jordanian twist of the XXX model with non-periodic boundary conditions was studied in \cite{KMN10}.
A way to introduce non-periodic boundary conditions compatible with the integrability of the quantum systems solvable by the QISM was developed in \cite{Sklyanin88}. The boundary conditions at the left and right end of the system are expressed in the left and right reflection matrices. The compatibility condition between the bulk and the boundary of the system takes the form of the so-called reflection equation \cite{Cherednik84,KulishSklyanin92}. The compatibility at the right end of the model is expressed by the dual reflection equation. The matrix form of the exchange relations between the entries of the Sklyanin monodromy matrix is analogous to the reflection equation. Together with the dual reflection equation they yield the commutativity of the open transfer matrix \cite{Sklyanin88,FreidelMaillet91}. The general solution of the reflection equation associated with the Jordanian deformation of the $SL(2)$-invariant Yang R-matrix was given in \cite{KMN10}, as well as the Hamiltonian of the deformed XXX spin chain with the general boundary terms.

Here, we study the deformation by the Jordanian twist of the $s\ell(2)$ Gaudin model with non-periodic boundary conditions. The starting point to obtain the Gaudin model in the framework of the QISM is the monodromy matrix of the corresponding inhomogeneous spin chain \cite{Sklyanin89}. The quasi-classical expansion of the  transfer matrix of the periodic chain, calculated at 
special values of the spectral parameter, produces the Gaudin Hamiltonians \cite{HikamiKulishWadati92,HikamiKulishWadati92a}. This approach was generalized to the case of non-periodic boundary conditions in \cite{Hikami95}. These results were later extended to non-diagonal reflection matrices \cite{YangZhangGould04} and to other simple Lie algebras \cite{YangZhangSasakic05}. Essential steps in the derivation of the
Gaudin Hamiltonians with boundary terms are some normalization conditions which have to be imposed both on the R-matrix and on the  left and right reflection matrices. In general, these normalization conditions are not relevant in the study of the spin chain. However, they enable the quasi-classical expansion of the transfer matrix which yields the corresponding Gaudin Hamiltonians with boundary terms \cite{YangZhangGould04,CAMN13}.

This paper is organized as follows. In Section II, the Jordanian deformation of the SL(2)-invariant Yang R-matrix is reviewed. The general solutions of the corresponding reflection equation and the dual reflection equation are given. In Section III, the Jordanian deformation of the inhomogeneous XXX spin chain with $N$ sites is studied. It is shown how the quasi-classical expansion of the transfer matrix yields the Gaudin Hamiltonians. These Hamiltonians are calculated in Section IV.
\section{Twisted Yang R-matrix and reflection equation}
We start this section by reviewing the Jordanian twist element. The three generators of the $s\ell(2)$ Lie algebra are $h, X^{\pm}$, such that
\begin{equation}
\label{sl2}
\left[ h, X^{\pm} \right]  = \pm 2 X^{\pm} , \quad \left[ X^+, X^- \right]   = \, h .
\end{equation}
The universal enveloping algebra $U(s\ell(2))$ admits the Jordanian twist element \cite{Gerstenhaber,Ogievetsky}
\begin{equation}
\label{jF}
\mathcal{F} = \exp (\frac{1}{2} h\otimes \ln (1+ 2 \theta X^+)) = e ^{h \otimes \sigma} \in U(s\ell(2)) \otimes U(s\ell(2))
\end{equation}
which satisfies the following equations
\begin{align}
\label{Jtwist1}
\left( \Delta \otimes \mathrm{id} \right) \left(  \mathcal{F} \right)   &=   \mathcal{F}_{13} \mathcal{F}_{23} ,\\
\label{Jtwist2}
\left(  \mathrm{id} \otimes \Delta_{t} \right) \left(  \mathcal{F} \right)   &=   \mathcal{F}_{12} \mathcal{F}_{13} .
\end{align}
The coproduct $\Delta$ is the usual coproduct of $U(s\ell(2))$ and $\Delta_{t}$ is the twisted one. In order to check the equations above, it is important to notice that $h$ is primitive with respect to $\Delta$ and that $\sigma$  is primitive with respect to $\Delta_{t}$,
\begin{align}
\Delta (h)  &= h \otimes 1 + 1 \otimes h,  \\
\Delta_{t} (\sigma) &= \sigma\otimes 1 + 1 \otimes \sigma .
\end{align}
The equations \eqref{Jtwist1} and  \eqref{Jtwist2} imply the twist equation \eqref{TwEq},
\begin{align}
\label{}
\mathcal{F}_{12} \left( \Delta \otimes \mathrm{id} \right) (\mathcal{F})  &= \mathcal{F}_{12} \mathcal{F}_{13} \mathcal{F}_{23} ,\\
\mathcal{F}_{23}\left(\mathrm{id} \otimes  \Delta\right) (\mathcal{F})    &=   \mathcal{F}_{23}\left(\mathrm{id} \otimes  \Delta\right) (\mathcal{F}) \mathcal{F}_{23}^{-1} \cdot \mathcal{F}_{23} = \left(  \mathrm{id} \otimes \Delta_{t} \right) \left(  \mathcal{F} \right) \mathcal{F}_{23} = \mathcal{F}_{12} \mathcal{F}_{13} \mathcal{F}_{23} .
\end{align}
Therefore the Jordanian twist element \eqref{jF} satisfies the Drinfeld twist equation \eqref{TwEq}.

The Heisenberg XXX spin chain is related to the Yangian $\mathcal{Y}(s\ell(2))$ and the $SL(2)$-invariant Yang R-matrix 
\begin{equation}
\label{YangR}
R (\lambda) = \left(\begin{array}{cccc}
1 & 0 & 0 & 0 \\[1ex]
0 & \displaystyle{\frac{\lambda}{\lambda + \eta}} & \displaystyle{\frac{\eta}{\lambda + \eta}} & 0 \\[1.5ex]
0 & \displaystyle{\frac{\eta}{\lambda + \eta}} & \displaystyle{\frac{\lambda}{\lambda + \eta}}  & 0 \\[1.5ex]
0 & 0 & 0 & 1\end{array}\right),
\end{equation}
where $\lambda$ is a spectral parameter, $\eta$ is a quasi-classical parameter.

The matrix form of the Jordanian twist element \eqref{jF} in the spin {\textonehalf}  representation $\rho$ is $F_{12}\in \mathrm{End} \left(\mathbb{C}^2 \otimes \mathbb{C}^2 \right)$
\begin{equation}
   \label{jFonehalf}
F_{12} = \left( \rho \otimes \rho \right) \mathcal{F} = \exp \left( \sigma ^z \otimes \theta \sigma ^+ \right) = \mathbbm{1} + \theta \sigma ^z \otimes \sigma ^+
     = \left(\begin{array}{rrrr}1 & \theta & 0 & 0 \\0 & 1 & 0 & 0 \\0 & 0 & 1 & -\theta \\0 & 0 & 0 & 1\end{array}\right) ,
\end{equation}
where $\sigma ^z,  \sigma ^{\pm} = (\sigma ^x \pm \imath \sigma ^y)/2$ are the Pauli sigma matrices. Hence, the twisted $R$-matrix has the form \cite{KulishStolin97,KMN10}
\begin{equation}
\label{Rmat}
R^{(J)}(\lambda)=F_{21}R_{12}(\lambda) F_{12}^{-1}  =
\left(\begin{array}{cccc}
1 & \displaystyle{\frac{-\lambda \theta}{\lambda + \eta}} & \displaystyle{\frac{\lambda \theta}{\lambda + \eta}} & \displaystyle{\frac{\lambda \theta^2}{\lambda + \eta}} \\[1.5ex]
0 & \displaystyle{\frac{\lambda}{\lambda + \eta}} & \displaystyle{\frac{\eta}{\lambda + \eta}} & \displaystyle{\frac{-\lambda \theta}{\lambda + \eta}}  \\[1.5ex]
0 & \displaystyle{\frac{\eta}{\lambda + \eta}} & \displaystyle{\frac{\lambda}{\lambda + \eta}}  & \displaystyle{\frac{\lambda \theta}{\lambda + \eta}}\\[1.5ex]
0 & 0 & 0 & 1
\end{array}\right) .
\end{equation}
This R-matrix is also a solution of the Yang-Baxter equation
\begin{equation}
\label{YBE}
R_{12} ( \lambda - \mu) R_{13} ( \lambda) R_{23} (\mu) = R_{23} (\mu ) R_{13} (\lambda ) R_{12} ( \lambda - \mu),
\end{equation}
we use the standard notation of the QISM  \cite{TakhFadI,KulishSklyanin82,Faddeev} to denote spaces $V_j, j=1, 2, 3$ on which corresponding $R$-matrices $R_{ij}, ij = 12, 13, 23$ act non-trivially. In the present case $V_1 = V_2 = V_3 =\mathbb{C}^2$. In what follows we will only use the twisted R-matrix \eqref{Rmat} and in order to simplify the notation we will drop the symbol $(J)$ in the superscript.

The Gaudin models are related to the classical r-matrices \cite{Sklyanin89}. Therefore it is essential that, after setting $\theta=-\xi \eta$,  the R-matrix has the quasi-classical property \cite{CAM05}
\begin{equation}
\label{Rsemiclass}
    R  (\lambda, \eta, \theta) |_{\theta=-\xi \eta} = \mathbbm{1} + \eta r(\lambda) + \mathcal{O} (\eta ^2),
\end{equation}
here $r(\lambda)$ is the corresponding classical r-matrix
\begin{equation}
\label{r-classical}
r(\lambda) = \left(\begin{array}{cccc}
0 & \xi & -\xi & 0 \\ [1ex]
0 & \displaystyle{\frac{-1}{\lambda}} & \displaystyle{\frac{1}{\lambda}} & \xi \\[1.5ex]
0 & \displaystyle{\frac{1}{\lambda}} & \displaystyle{\frac{-1}{\lambda}} & -\xi \\[1.5ex]
0 & 0 & 0 & 0\end{array}\right),
\end{equation}
which has the unitarity property
\begin{equation}
\label{r-unitarity}
r_{21}(-\lambda) = - r_{12}(\lambda) ,
\end{equation}
and satisfies the classical Yang-Baxter equation
\begin{equation}
\label{classicalYBE}
[r_{13} (\lambda), r_{23}(\mu) ] + [r_{12}(\lambda - \mu), r_{13} (\lambda) +  r_{23}(\mu)] =0.
\end{equation}
Moreover, for the purpose of deriving the Gaudin Hamiltonians, it is necessary that
the R-matrix \eqref{Rmat} is normalized so that \cite{HikamiKulishWadati92, HikamiKulishWadati92a}
\begin{equation}
\label{Rlambdazero}
    R (0, \eta) = \mathcal{P},
\end{equation}
where $\mathcal{P}$ is the permutation matrix in $\mathbb{C}^2 \otimes \mathbb{C}^2$. The R-matrix \eqref{Rmat} has the unitarity property
\begin{equation}
\label{unitarity}
R_{12} ( \lambda) R_{21} ( - \lambda) = \mathbbm{1},
\end{equation}
but the PT symmetry is broken
\begin{equation}
R_{21}(\lambda)\neq R_{12}(\lambda)^{t_1t_2},
\end{equation}
where $R_{21}(\lambda) = \mathcal{P} R_{12}(\lambda) \mathcal{P}$, and the indices $t_1$ and $t_2$ denote the respective transpositions in the first and second space of the tensor product $\mathbb{C}^2 \otimes \mathbb{C}^2$.
The $R$-matrix also does not have the crossing symmetry, but it dos satisfy the weaker condition
\begin{equation}
\label{weakcross}
\{\{\{R_{12}(\lambda)^{t_2}\}^{-1}\}^{t_2}\}^{-1}= g(\lambda) M_2 R_{12}(\lambda+2\eta)M_2^{-1} ,
\end{equation}
with $g(\lambda) = \displaystyle{\frac{\lambda (\lambda + 2 \eta)}{(\lambda + \eta) ^2}}$ and the matrix
\begin{equation}
\label{Mmat}
M=\left(\begin{array}{cc} 1 & -2\theta\\ 0 & 1 \end{array} \right) .
\end{equation}
We note that a more general matrix
\begin{equation}
\label{Mgen}
\widetilde{M}=\left(\begin{array}{cc} 1 & \alpha \\ 0 & 1 \end{array} \right)
\end{equation}
commutes with the $R$-matrix,
\begin{equation}
\label{Mcom}
\left[ \widetilde{M} \otimes \widetilde{M} , R(\lambda) \right]=0 .
\end{equation}

A way to introduce non-periodic boundary conditions compatible with the integrability of the base model was developed in \cite{Sklyanin88}. The boundary conditions at the left and right sites of the system are expressed in the left and right reflection matrices $K^-$ and $K^+$. The compatibility condition between the bulk and the boundary of the system takes the form of the so-called reflection equation \cite{Cherednik84,KulishSklyanin92}. For the left reflection matrix, it is written in the form
\begin{equation}
\label{RE}
R_{12}(\lambda - \mu) K^-_1(\lambda) R_{21}(\lambda + \mu) K^-_2(\mu)=
K^-_2(\mu) R_{12}(\lambda + \mu) K^-_1(\lambda) R_{21}(\lambda - \mu) .
\end{equation}
The compatibility at the right site of the model is expressed by the dual reflection equation \cite{Sklyanin88,FreidelMaillet91,KMN10}
\begin{equation}
\label{rightRE}
A_{12}(\lambda - \mu) K^{+\, t}_1(\lambda) B_{12}(\lambda + \mu) K^{+\, t}_{2}(\mu) =
K_2^{+\, t}(\mu) C_{12} (\lambda + \mu) K_1^{+\, t}(\lambda) D_{12}(\lambda - \mu) .
\end{equation}
where the matrices $A,B,C,D$ are obtained from the $R$-matrix of reflection equation \eqref{RE} as
\begin{align}
A_{12}(\lambda)&=\left(R_{12}(\lambda)^{t_{12}}\right)^{-1}=D_{21}(\lambda) , \\
B_{12}(\lambda)&= \left( \left(R_{21}^{t_1}(\lambda)\right)^{-1}\right)^{t_2}=C_{21}(\lambda) .
\end{align}
Using property \eqref{weakcross}, we can write dual reflection equation \eqref{rightRE} in the equivalent form
\begin{align}
\label{dRE}
&R_{12}( -\lambda + \mu )K_1^{+}(\lambda) M_2 R_{21}(-\lambda - \mu - 2\eta) M_2^{-1} K_2^{+}(\mu)= \notag\\
&K_2^{+}(\mu) M_1 R_{12}(-\lambda -\mu-2\eta) M_1^{-1} K_1^{+}(\lambda) R_{21}(-\lambda + \mu) .
\end{align}
It can then be verified that the mapping
\begin{equation}
\label{bijectionKpl}
K^+(\lambda)= K^{-}(- \lambda -\eta) \ M
\end{equation}
is a bijection between solutions of the reflection equation and the dual reflection equation.

The general solution of the reflection equation associated with the Jordanian deformation of the Yang R-matrix is \cite{KMN10}
\begin{equation}
\label{Kminus}
K^{-} (\lambda) = \frac{1}{d (\lambda)} \left(\begin{array}{cc}
\zeta -\lambda - \displaystyle{\frac{\phi\theta}{\eta}} \lambda ^2 & \psi \lambda  \\[1.5ex]
\phi \lambda & \zeta + \lambda - \displaystyle{\frac{\phi\theta}{\eta}} \lambda ^2
\end{array}\right),
\end{equation}
with arbitrary parameters $\zeta$, $\phi$ and $\psi$ and the function
\begin{equation}
\label{d-function}
d (\lambda) = \zeta -\displaystyle{\frac{\phi\theta}{\eta}} \lambda ^2 + \lambda \sqrt{1+\phi\psi} \ .
\end{equation}
The left reflection matrix $K^-$ is normalized so that
\begin{equation}
\label{normalizationK}
K^{-} (\lambda) K^{-} (- \lambda) = \mathbbm{1}.
\end{equation}
Finally, by setting $\theta=-\xi \eta$ we achive that matrix $K^{-} (\lambda)$ does not depend on the quasi-classical parameter $\eta$,
\begin{equation}
\label{etaKminus}
\frac{\partial K^{-} (\lambda)}{\partial \eta} = 0.
\end{equation}
The right reflection matrix $K^+(\lambda)$ is obtained by substituting \eqref{Kminus} into \eqref{bijectionKpl} while keeping
$\theta=-\xi \eta$. It is important to notice that
\begin{equation}
\label{normalizationKpl}
\left( \lim_{\eta \to 0} K^+(\lambda) \right) K^{-} (\lambda) = \mathbbm{1}.
\end{equation}
In general, the normalization conditions \eqref{normalizationK}  and \eqref{normalizationKpl} are not essential in  the study of the open spin chain. However, together with \eqref{Rsemiclass} and \eqref{Rlambdazero} they enable the quasi-classical expansion of the transfer matrix which yields the open Gaudin model.

\section{Deformed inhomogeneous XXX chain}

In this section we study the Jordanian deformation of the inhomogeneous XXX spin chain with $N$ sites, characterized by the local space $V_j= \mathbb{C}^2$ and inhomogeneous parameter $\alpha _j$. For simplicity, we start by considering the periodic boundary conditions. The Hilbert space of the system is
$$
\mathcal{H} = \underset {j=1}{\overset {N}{\otimes}}  V_j = (\mathbb{C}^2 ) ^{\otimes N}.
$$
In the QISM \cite{TakhFadI, Faddeev,Kulish09} the so-called monodromy matrix
\begin{equation}
\label{monodromy-T}
T(\lambda ) = R_{0N} ( \lambda - \alpha _N) \cdots R_{01} ( \lambda - \alpha _1)
\end{equation}
is used to describe the system. For simplicity we have omitted the dependence on the quasi-classical parameter $\eta$, the deformation parameter $\xi$ and the inhomogeneous parameters $\{ \alpha _j , j = 1 , \ldots , N \}$. Notice that $T(\lambda)$ is a two-by-two matrix in the auxiliary space $V_0 = \mathbb{C}^2$, whose entries are operators acting in $\mathcal{H}$. Due to the Yang-Baxter equation \eqref{YBE}, it is straightforward to check that the monodromy matrix satisfies the RTT-relations \cite{TakhFadI, Faddeev,KulishSklyanin82}
\begin{equation}
\label{RTT}
R_{12} ( \lambda - \mu) \underset {1} {T}(\lambda ) \underset {2} {T}(\mu ) = \underset {2} {T}(\mu )\underset {1} {T}(\lambda ) R_{12} ( \lambda - \mu).
\end{equation}
The above equation is written in the tensor product of the auxiliary space $V_0\otimes V_0 = \mathbb{C}^2 \otimes \mathbb{C}^2$, using the standard notation of the QISM.

The periodic boundary conditions and the RTT-relations \eqref{RTT} imply that the transfer matrix
\begin{equation}
\label{periodic-t}
t (\lambda ) = \mathrm{tr}_0 T(\lambda) ,
\end{equation}
commute at different values of the spectral parameter,
\begin{equation}
\label{periodic-tt}
[t (\lambda) , t (\mu)] = 0,
\end{equation}
here we have omitted the nonessential arguments.

Due to  the quasi-classical property \eqref{Rsemiclass} and the normalization of the R-matrix \eqref{Rlambdazero}, the quasi-classical expansion of the transfer matrix, for the special values of the spectral parameter, is given by \cite{HikamiKulishWadati92,HikamiKulishWadati92a}
\begin{equation}
\label{periodic-Zk}
Z_k = t (\lambda = \alpha _k ) = \mathbbm{1} + \eta H_k + \mathcal{O} (\eta ^2),
\end{equation}
where $H_k$ are the corresponding Gaudin Hamiltonians, in the periodic case. The commutativity of the Gaudin Hamiltonians,
\begin{equation}
\label{commHk}
[H_k, H_l] =0,
\end{equation}
is ensured by the commutativity of the transfer matrix \eqref{periodic-tt} and the fact the first term in the above expansion is the identity matrix, due to \eqref{Rsemiclass} and \eqref{Rlambdazero}.

In order to construct integrable spin chains with non-periodic boundary condition, we use the Sklyanin formalism \cite{Sklyanin88}. The corresponding monodromy matrix $\mathcal{T}(\lambda)$ consists of the two matrices $T(\lambda)$ \eqref{monodromy-T} and a reflection matrix $K ^{-}(\lambda)$ \eqref{Kminus},
\begin{align}
\label{double-T}
\mathcal{T}(\lambda) &= T(\lambda) K ^{-}(\lambda) T^{-1}(- \lambda)  \notag \\
    &= R_{0N}( \lambda - \alpha _N) \cdots R_{01}(\lambda - \alpha _1) K _0^{-}(\lambda) R_{10}(\lambda + \alpha _1 )
\cdots R_{N0}(\lambda + \alpha _N ),
\end{align}
where, for simplicity, we have suppressed the dependence on the other parameters. By construction, the exchange relations of the monodromy matrix $\mathcal{T}(\lambda)$  in $V_0\otimes V_0$ are
\begin{equation}
\label{exchangeRE}
R_{12}(\lambda - \mu) \underset {1} {\mathcal{T}}(\lambda) R_{21}(\lambda + \mu) \underset {2} {\mathcal{T}}(\mu)=
\underset {2} {\mathcal{T}}(\mu) R_{12}(\lambda + \mu) \underset {1}{\mathcal{T}}(\lambda) R_{21}(\lambda - \mu) ,
\end{equation}
using the notation of \cite{Sklyanin88}. The open chain transfer matrix is given by the trace of $\mathcal{T}(\lambda)$ over the auxiliary space $V_0$ with an extra reflection matrix $K^+(\lambda)$ \cite{Sklyanin88},
\begin{equation}
\label{open-t}
t (\lambda) = \mathrm{tr}_0 \left( K^+(\lambda) \mathcal{T}(\lambda) \right).
\end{equation}
The reflection matrix $K^+(\lambda)$ \eqref{bijectionKpl} is the corresponding solution of the dual reflection equation \eqref{dRE}. The commutativity of the transfer matrix for different values of the spectral parameter
\begin{equation}
\label{open-tt}
[t (\lambda) , t (\mu)] = 0,
\end{equation}
is guaranteed by the dual reflection equation \eqref{dRE} and the exchange relations \eqref{exchangeRE} of the monodromy matrix $\mathcal{T}(\lambda)$.

Analogously to the periodic case, the quasi-classical expansion of the transfer matrix, for special values of the spectral parameter, yields the corresponding Gaudin Hamiltonians with boundary terms \cite{YangZhangGould04}
\begin{equation}
\label{open-Zk}
Z_k = t (\lambda = \alpha _k ) = \mathbbm{1} + \eta H_k + \mathcal{O} (\eta ^2),
\end{equation}
where $H_k$ are the Gaudin Hamiltonians.  As in the periodic case, the commutativity of the Hamiltonians $H_k$ is guaranteed by the equation \eqref{open-tt} and the fact the first term in the above expansion is the identity matrix, due to the quasi-classical property of the R-matrix \eqref{Rsemiclass} and the normalization conditions \eqref{Rlambdazero}, \eqref{normalizationK} and \eqref{normalizationKpl}. The Gaudin Hamiltonians will be calculated in the following section.

\section{Deformed Gaudin Hamiltonians with boundary terms}

In this section we calculate explicitly the Gaudin Hamiltonians.  In the periodic case \cite{HikamiKulishWadati92, HikamiKulishWadati92a}, it is straightforward to calculate the first two terms in the expansion \eqref{periodic-Zk}. By setting $\theta=-\xi \eta$ and $\lambda = \alpha _k$ and using the normalization of the R-matrix \eqref{Rlambdazero} we obtain
\begin{align}
\label{firstperiodic-Zk}
Z_k |_{\eta = 0} &= \mathrm{tr}_0 \left( R_{0N} ( \alpha _k - \alpha _N) \cdots R_{0k} ( \alpha _k - \alpha _k) \cdots R_{01} ( \alpha _k - \alpha _1) \right) |_{\eta = 0}  \notag \\
    &= \mathrm{tr}_0 \left( R_{0N} ( \alpha _k - \alpha _N) \cdots \mathcal{P}_{0k} \cdots R_{01} ( \alpha _k - \alpha _1) \right) |_{\eta = 0} \notag \\
    &= \mathrm{tr}_0 \left( \mathcal{P}_{0k} \right) = \mathbbm{1}.
\end{align}
In order to derive the corresponding Gaudin Hamiltonians, in the second term we also use the quasi-classical property of the R-matrix \eqref{Rsemiclass},
\begin{align}
\label{GH-periodic}
H_k = \frac{\partial Z_k}{\partial \eta} |_{\eta = 0} &= \sum _{l > k }\mathrm{tr}_0 \left( R_{0N} ( \alpha _k - \alpha _N) \cdots \frac{\partial R_{0l}( \alpha _k - \alpha _l)}{\partial \eta} \cdots \mathcal{P}_{0k} \cdots R_{01} ( \alpha _k - \alpha _1) \right) |_{\eta = 0} \notag \\
&+ \sum _{l < k }\mathrm{tr}_0 \left( R_{0N} ( \alpha _k - \alpha _N) \cdots \mathcal{P}_{0k} \cdots \frac{\partial R_{0l}( \alpha _k - \alpha _l)}{\partial \eta} \cdots R_{01} ( \alpha _k - \alpha _1) \right) |_{\eta = 0} \notag \\
 &= \sum _{l > k } \mathrm{tr}_0 \left(  \frac{\partial R_{0l}( \alpha _k - \alpha _l)}{\partial \eta} |_{\eta = 0}\mathcal{P}_{0k} \right) + \sum _{l < k }\mathrm{tr}_0 \left(  \mathcal{P}_{0k} \frac{\partial R_{0l}( \alpha _k - \alpha _l)}{\partial \eta} |_{\eta = 0} \right) \notag \\
 &= \sum _{l \neq k } r_{kl}( \alpha _k - \alpha _l).
\end{align}%
In terms of the local spin-\textonehalf \ operators $\overrightarrow{S_j} = \displaystyle{\frac{\overrightarrow{\sigma_j}}{2}}$ these Hamiltonians are
\begin{equation}
\label{GHam-periodic}
H_k \simeq  2 \sum _{l \neq k }  \left( \displaystyle{\frac{\overrightarrow{S_k} \cdot \overrightarrow{S_l}}{\alpha_k - \alpha_l}} + \xi \left( S^z_k S^+_l - S^+_l S^z_k \right) \right),
\end{equation}
where $\simeq$ stands for equality up to a constant additive term. By implementation of the algebraic Bethe ansatz, it was shown in \cite{CAM05} that the spectra of these Hamiltonians coincide with the spectra of the $s\ell(2)$ invariant model, the case when $\xi = 0$.

In the case of non-periodic boundary, we consider the left and right reflection matrices $K^{-}(\lambda)$ and $K^{+}(\lambda)$  given by \eqref{Kminus} and \eqref{bijectionKpl}, respectively, together with $\theta=-\xi \eta$. In order to obtain the expansion \eqref{open-Zk}
in \eqref{open-t} we specify $\lambda = \alpha _k$ and calculate the first term
\begin{align}
\label{firstopen-Zk}
Z_k |_{\eta = 0} &= \mathrm{tr}_0 \left( K^+(\alpha _k) \mathcal{T}(\alpha _k) \right) |_{\eta = 0}
= \mathrm{tr}_0 \left( K_0^+(\alpha _k)  R_{0N} ( \alpha _k - \alpha _N) \cdots \mathcal{P}_{0k} \cdots R_{01} ( \alpha _k - \alpha _1)  \right. \times \notag \\
&\left. \times K _0^{-}(\alpha _k) R_{10}(\alpha _k + \alpha _1 ) \cdots R_{k0}( 2\alpha _k ) \cdots R_{N0}(\alpha _k + \alpha _N ) \right) |_{\eta = 0} \notag \\
&= \mathrm{tr}_0 \left( K_0^-( - \alpha _k) \mathcal{P}_{0k} K _0^{-}(\alpha _k) \right) =  K _k^{-}(\alpha _k)K_k^-( - \alpha _k) =\mathbbm{1}.
\end{align}
In the last step above we have used the normalization \eqref{normalizationK}. The Gaudin Hamiltonians with boundary terms are related to the second term in the expansion \eqref{open-Zk},
\begin{align}
\label{calc-GH-boundary}
H_k = \frac{\partial Z_k}{\partial \eta} |_{\eta = 0} &= \mathrm{tr}_0 \left( \frac{\partial K_0^+(\alpha _k)}{\partial \eta} |_{\eta = 0} \mathcal{P}_{0k}  K _0^{-}(\alpha _k) \right) + \mathrm{tr}_0 \left(  K_0^-( - \alpha _k) \mathcal{P}_{0k} K _0^{-}(\alpha _k)  \frac{\partial R_{k0}( 2 \alpha _k )}{\partial \eta} |_{\eta = 0} \right) \notag \\
&+ \sum _{l > k }\mathrm{tr}_0 \left( K_0^-( - \alpha _k)  \frac{\partial R_{0l}( \alpha _k - \alpha _l)}{\partial \eta} |_{\eta = 0} \mathcal{P}_{0k} K _0^{-}(\alpha _k) \right) \notag \\
&+ \sum _{l < k }\mathrm{tr}_0 \left( K_0^-( - \alpha _k)  \mathcal{P}_{0k}  \frac{\partial R_{0l}( \alpha _k - \alpha _l)}{\partial \eta} |_{\eta = 0} K _0^{-}(\alpha _k) \right) \notag \\
&+ \sum _{l \neq k }\mathrm{tr}_0 \left( K_0^-( - \alpha _k)  \mathcal{P}_{0k} K _0^{-}(\alpha _k) \frac{\partial R_{l0}( \alpha _k + \alpha _l)}{\partial \eta} |_{\eta = 0} \right) .
\end{align}
In the derivation above we have used the fact that the left reflection matrix $ K^{-}( \lambda)$ does not depend  on the quasi-classical parameter $\eta$, \eqref{etaKminus}. Finally, the Gaudin Hamiltonians can be expressed in a more concise form
\begin{equation}
\label{GH-boundary}
H_k = \Gamma_k (\alpha _k ) + \sum _{l \neq k } r_{kl}( \alpha _k - \alpha _l) + \sum _{l \neq k } K _k^{-}(\alpha _k) r_{lk}( \alpha _k + \alpha _l)   K_k^-( - \alpha _k),
\end{equation}
where $r_{ij}(\lambda)$ is the corresponding classical r-matrix \eqref{r-classical} and
\begin{equation}
\label{Gamma-k}
\Gamma_k (\alpha _k ) = K _k^{-}(\alpha _k) \left( \frac{\partial K_k^+(\alpha _k)}{\partial \eta} |_{\eta = 0} + \mathrm{tr}_0 \left(  K_0^-( - \alpha _k) \ \mathcal{P}_{0k} \ r_{{k0}}(2 \alpha _k) \right) \right) .
\end{equation}
Notice that the second term on the righthand side of the \eqref{GH-boundary} coincides with the Gaudin Hamiltonians \eqref{GH-periodic} and that the first and the third term are the boundary terms depending on the reflection matrices $K^-$ and $K^+$. Consequently, the Hamiltonians \eqref{GH-boundary} are rational functions of the boundary parameters $\zeta, \phi, \psi$ and the deformation parameter $\xi$.

\section{Conclusions}

We have derived the deformed $s\ell(2)$ Gaudin Hamiltonians with boundary terms. Starting from the $SL(2)$-invariant Yang R-matrix deformed by the Jordanian twist element for the universal enveloping algebra $U(s\ell(2))$ and generic solutions of the associated reflection equation and the dual reflection equation, we have obtained the deformed inhomogeneous spin-\textonehalf \ XXX chain. We have shown that the quasi-classical expansion of the transfer matrix of the chain yields the corresponding Gaudin Hamiltonians with boundary terms.

\bigskip

\textbf{Acknowledgments.} This work was supported by the FCT Project No. \hfil \break PTDC/MAT/099880/2008 through the European program COMPETE/FEDER.

\end{document}